\documentclass[prl,notitlepage,nofootinbib,twocolumn]{revtex4-1}

\usepackage{amsmath}

\usepackage{float}
\usepackage{graphicx}	
\usepackage{hyperref}
\usepackage{mathtools}
\usepackage{cancel}
\usepackage{multirow}
\usepackage{rotating}
\usepackage{physics}

\begin{document}

\title{Interference between the Atmospheric and Solar Oscillation Amplitudes}

\author{Patrick Huber}
\email{pahuber@vt.edu}
\author{Hisakazu Minakata}
\email{minakata71@vt.edu}
\author{Rebekah Pestes}
\email{rebhawk8@vt.edu}
\affiliation{Center for Neutrino Physics, Department of Physics, Virginia Tech, Blacksburg, Virginia 24061, USA}

\date{\today}
\begin{abstract}
We propose to detect the interference effect between the
atmospheric-scale and solar-scale waves of neutrino oscillation, one
of the key consequences of the three-generation structure of
leptons. In vacuum, we show that there is a natural and general way of
decomposing the oscillation amplitude into these two oscillation
modes. The nature of the interference is cleanest in the $\bar{\nu}_e$
disappearance channel since it is free from the CP-phase $\delta$. We
find that the upcoming JUNO experiment offers an ideal setting to
observe this interference with more than $4\,\sigma$ significance,
even under conservative assumptions about the systematic
uncertainties.  Finally, we discuss the relationship between the mass
ordering resolution and the interference effect.
\end{abstract} 

\maketitle

\section{Introduction}
\label{sec:introduction } 

It is a remarkable feature of nature that the fundamental fermions,
quarks, and leptons come into our world in the form of three
generations, which has various important consequences. The most
dramatic one among them would be to provide a mechanism for CP
violation~\cite{Christenson:1964fg}. 
The Standard Model of particle physics with three families of quarks, not
two, allows for the existence of a phase, the Kobayashi-Maskawa (KM)
phase~\cite{Kobayashi:1973fv}, by which CP symmetry is broken; and
indeed, CP violation has been observed~\cite{Aubert:2001nu,Abe:2001xe,Tanabashi:2018oca}.
If a similar phase exists in the lepton sector in the neutrino-mass-embedded Standard Model ($\nu$SM), then there will be CP violation 
due to the lepton KM phase.\footnote{
Note, if neutrinos are Majorana particles~\cite{Majorana:1937vz}, there
is an option of having CP violation with only two generations of
leptons.  }
The observation of the leptonic
CP phase is being and will be actively pursued by the ongoing and
next generation neutrino oscillation experiments~\cite{Abe:2019vii,Acero:2019ksn,Jiang:2019xwn,Abe:2015zbg,Acciarri:2015uup}.

It is a natural question whether CP violation is the only consequence
of the three family structure. In the context of neutrino oscillation,
the three generation structure allows for the existence of two
independent mass squared differences: $\Delta m^2_{31} \equiv m_{3}^2
- m_{1}^2$ and $\Delta m^2_{21}$. Experimentally, we find the $\Delta
m^2_{31}$-driven atmospheric neutrino oscillation
\cite{Fukuda:1998mi} and the solar $\Delta m^2_{21}$-driven reactor
neutrino oscillation~\cite{Eguchi:2002dm}, as well as the $\Delta
m^2_{21}$-matter potential induced flavor conversion~\cite{Wolfenstein:1977ue,Mikheev:1986gs} inside the sun~\cite{Ahmad:2002jz}.

Building on this success, in this paper, we wish to add a new item to
the list of nontrivial consequences of the three generation
structure: Quantum interference between the atmospheric-scale and
solar-scale waves of neutrino oscillation.  So far, the existence of
the small $\Delta m^2_{21}$ effects in atmospheric and long-baseline
(LBL) accelerator neutrino experiments and, similarly, the effects of
the larger $\Delta m^2_{31}$, as well as the $\theta_{13}$ mixing
effect, in the solar neutrino observation have been recognized as small
sub-leading effects.
The simultaneous full existence of both the $\Delta m^2_{31}$ and
$\Delta m^2_{21}$ waves and their mutual interference, if observed,
would establish another consequence of the three generation structure
of neutrinos embedded into the $\nu$SM.
For previous discussions which addressed related interference phenomena, see {\it e.g.}, Refs.~\cite{Smirnov:2006sm,Akhmedov:2008qt} for accelerator/atmospheric neutrinos, and \cite{Petcov:2001sy,Choubey:2003qx,Learned:2006wy} for reactor neutrinos.

\section{The atmospheric and solar amplitudes} 
\label{sec:Atm-solar-amplitudes}

Our first task is to define what the atmospheric and solar amplitudes
are in neutrino oscillation. In this paper, we restrict our
discussion to vacuum, as a similar generic definition fulfilling the conditions 1 and 2 below is not available --- in fact, very likely not existing --- in matter
\cite{HMP-matter}.\footnote{
The authors of ref.~\cite{Akhmedov:2008qt} propose a particular way of
decomposition into the ``atmospheric'' and ``solar'' amplitudes in
matter. We will contrast their method to our own proposal in
ref.~\cite{HMP-matter}. }
The flavor basis $S$ matrix elements $S_{\alpha \beta}$ ($\alpha,
\beta = e, \mu, \tau$), which describe the neutrino flavor
transformation $\nu_{\beta} \rightarrow \nu_{\alpha}$, can be written
under the ultra-relativistic approximation of neutrinos as
\begin{eqnarray} 
S_{\alpha \beta} &=& 
U_{\alpha 1} U^{*}_{\beta 1} 
+ U_{\alpha 2} U^{*}_{\beta 2} e^{ - i \frac{ \Delta m^2_{21} }{2 E} x} 
+ U_{\alpha 3} U^{*}_{\beta 3} e^{ - i \frac{ \Delta m^2_{31} }{2 E} x},
\hspace{3mm}
\label{S-matrix}
\end{eqnarray}
where $E$ is the energy and $\Delta m^2_{ji} \equiv m^2_{j} - m^2_{i}$
$(i, j=1,2,3)$ denote the mass squared differences of
neutrinos. $U_{\alpha i}$ is the element of the lepton flavor mixing
matrix which relates the flavor and the mass eigenstates of neutrino
as $\nu_{\alpha} = U_{\alpha i}\nu_{i}$. In Eq.~\eqref{S-matrix}, we
factor out $e^{ - i m^2_{1} x / 2E}$ for simplicity of the expression,
which of course does not alter the physical observables. The
oscillation probability of the process $\nu_{\beta} \rightarrow
\nu_{\alpha}$ is given by $P(\nu_{\beta} \rightarrow \nu_{\alpha}:x) =
\vert S_{\alpha \beta} \vert^2$.
Hereafter, again for simplicity of the expressions, we define
\begin{eqnarray}
\Delta_{ji} \equiv \frac{\Delta m_{ji}^2}{2 E}.
\end{eqnarray}


We take a heuristic way to find the appropriate definitions of the
atmospheric and solar amplitudes. Let us first discuss the
appearance channel, $\alpha \neq \beta$. The $S$ matrix elements in
Eq.~\eqref{S-matrix} can be rewritten as
\begin{eqnarray} 
S_{\alpha \beta} = 
U_{\alpha 3} U^{*}_{\beta 3} \left( e^{ - i \Delta_{31} x} - 1 \right) 
+ U_{\alpha 2} U^{*}_{\beta 2}  \left( e^{ - i \Delta_{21} x} -1 \right)
\hspace{3mm}
\label{S-matrix-1}
\end{eqnarray}
due to unitarity, 
$U_{\alpha 1} U^{*}_{\beta 1} + U_{\alpha 2} U^{*}_{\beta 2} + U_{\alpha 3} U^{*}_{\beta 3} =0$.
Then, we claim that 
\begin{eqnarray} 
S_{\alpha \beta}^{ \text{atm} } \equiv  
U_{\alpha 3} U^{*}_{\beta 3} 
\left( e^{ - i \Delta_{31} x} - 1 \right)
\label{atm-amplitude-def}
\end{eqnarray}
is the atmospheric amplitude, and 
\begin{eqnarray} 
S_{\alpha \beta}^{ \text{sol} } \equiv  
U_{\alpha 2} U^{*}_{\beta 2}  \left( e^{ - i \Delta_{21} x} -1 \right)
\label{sol-amplitude-def}
\end{eqnarray}
is the solar amplitude. The atmospheric amplitude, by definition,
describes neutrino oscillation due to non-vanishing $\Delta m^2_{31}$,
and the solar amplitude describes the one caused by $\Delta
m^2_{21}$. Therefore, the obtained expressions \eqref{atm-amplitude-def} and
\eqref{sol-amplitude-def} for them are entirely natural ones.

In disappearance channels, due to a difference in unitarity, $U_{\alpha
  1} U^{*}_{\alpha 1} + U_{\alpha 2} U^{*}_{\alpha 2} + U_{\alpha 3}
U^{*}_{\alpha 3} =1$, the $S$ matrix has a slightly different
expression when it is written in terms of the atmospheric and the
solar amplitudes,
\begin{eqnarray} 
S_{\alpha \alpha} 
&=& 
1 + \vert U_{\alpha 3} \vert^2 \left( e^{ - i \Delta_{31} x} - 1 \right) 
+ \vert U_{\alpha 2} \vert^2 \left( e^{ - i \Delta_{21} x} -1 \right) 
\nonumber \\ 
&=& 
1 + S_{\alpha \alpha}^{ \text{atm} } 
+ S_{\alpha \alpha}^{ \text{sol} } 
\label{S-matrix-disapp}
\end{eqnarray}
where $S_{\alpha \alpha}^{ \text{atm} }$ and $S_{\alpha \alpha}^{
  \text{sol} }$ are defined by extending the definition in \eqref{atm-amplitude-def} and \eqref{sol-amplitude-def}, by setting
$\beta = \alpha$. They, of course, satisfy the conditions $S_{\alpha
  \alpha}^{ \text{atm} } \rightarrow 0$ when $\Delta_{31} \rightarrow
0$, and $S_{\alpha \alpha}^{ \text{sol} } \rightarrow 0$ when
$\Delta_{21} \rightarrow 0$, respectively.

Now, we try to elevate the heuristic definitions into the general
definition of $S_{\alpha \beta}^{ \text{atm} }$ and $S_{\alpha
  \beta}^{ \text{sol} }$. For a given $S$ matrix element $S_{\alpha \beta}$
\begin{enumerate} 
\item
The atmospheric and the solar amplitudes are defined, respectively, as
\begin{eqnarray} 
S_{\alpha \beta}^{ \text{atm} } 
= \lim_{\Delta m^2_{21} \rightarrow 0} S_{\alpha \beta}, 
\hspace{6mm}
S_{\alpha \beta}^{ \text{sol} } 
= \lim_{\Delta m^2_{31} \rightarrow 0} S_{\alpha \beta}.
\label{atm-sol-amplitude-def}
\end{eqnarray}
\item 
We demand the completeness condition \\
$S_{\alpha \beta} = 
\delta_{\alpha \beta} + S_{\alpha \beta}^{ \text{atm} } + S_{\alpha \beta}^{ \text{sol} }$. 
\end{enumerate}
where $\delta_{\alpha \beta}$ denotes the Kronecker delta function.
Consistency requires the so obtained amplitudes to satisfy 
$\lim_{\Delta m^2_{31} \rightarrow 0} S_{\alpha \beta}^{ \text{atm} } 
= \lim_{\Delta m^2_{21} \rightarrow 0} S_{\alpha \beta}^{ \text{sol} } = 0$. 

The second condition, the completeness condition, demands that
decomposition of the oscillation amplitude into the atmospheric and
solar amplitudes is complete. We only have three neutrino
states and, therefore, two independent $\Delta m^2$, the atmospheric
$\Delta m^2_{31}$ and the solar $\Delta m^2_{21}$. So, there should
be two independent amplitudes, not more, not less.

\section{$\nu_{\mu} \rightarrow \nu_{e}$ and $\nu_{e} \rightarrow \nu_{e}$ channels} 
\label{sec:mue-ee-channels}

To obtain a sense of what the atmospheric and solar amplitudes
are, we write down their explicit forms in the $\nu_{\mu} \rightarrow
\nu_{e}$ and $\nu_{e} \rightarrow \nu_{e}$ channels by using the
flavor mixing matrix using the Particle Data Group (PDG) convention~\cite{Tanabashi:2018oca}. We leave the discussions of the other
channels to ref.~\cite{HMP-matter}.

The atmospheric and solar amplitudes, as defined in
Eqs.~\eqref{atm-amplitude-def} and \eqref{sol-amplitude-def},
respectively, can be written in the $\nu_{\mu} \rightarrow \nu_{e}$
channel as
\begin{eqnarray} 
&& S_{e \mu}^{ \text{atm} } = 
2i s_{23} c_{13} s_{13} 
e^{-i\delta} e^{ - i \frac{ \Delta_{31} x }{2} } 
\sin \frac{ \Delta_{31} x }{2}, 
\\
&& S_{e \mu}^{ \text{sol} } = 
2i s_{12} c_{13}
e^{ - i \frac{ \Delta_{21} x }{2} }
\left( c_{12} c_{23} - s_{12} s_{23} s_{13} e^{ - i\delta} \right) 
\sin \frac{ \Delta_{21} x }{2}
.
\nonumber
\label{atm-sol-amplitude-emu} 
\end{eqnarray}
The oscillation probability consists of two terms, each amplitude squared and summed and the interference term:
\begin{eqnarray} 
&&
P(\nu_{\mu} \rightarrow \nu_{e}) 
= \vert S_{e \mu}^{ \text{atm} } + S_{e \mu}^{ \text{sol} } \vert^2
\equiv P_{\mu e}^{ \text{non-int-fer} }
+ P_{\mu e}^{ \text{int-fer} },
\hspace{4mm}
\label{probability-decomposed-mue} 
\end{eqnarray}
where 
\begin{eqnarray} 
&&
P_{\mu e}^{ \text{non-int-fer} } \equiv
\vert S_{e \mu}^{ \text{atm} } \vert^2 
+ \vert S_{e \mu}^{ \text{sol} } \vert^2 
\nonumber \\
&&=
s^2_{23} \sin^2 2 \theta_{13} \sin^2 \frac{ \Delta_{31} x }{2}
+\sin^2 \frac{ \Delta_{21} x }{2}
\nonumber \\
&&\times
\biggl[ 
c^2_{23} c^2_{13} \sin^2 2\theta_{12} 
+ s^2_{23} s^4_{12} \sin^2 2\theta_{13} 
- 8 s^2_{12} J_r \cos \delta 
\biggr] 
, 
\nonumber \\
&&
P_{\mu e}^{ \text{int-fer} } \equiv 
2 \mbox{Re} \left[ \left( S_{e \mu}^{ \text{atm} } \right)^* S_{e \mu}^{ \text{sol} } \right]
\nonumber \\ 
&&=
8 \left[ J_r \cos \left( \delta + \frac{ \Delta_{32} x }{2} \right) 
- s^2_{23} c^2_{13} s^2_{13} s^2_{12} 
\cos \left( \frac{ \Delta_{32} x }{2} \right) 
\right]
\nonumber \\ 
&&\times
\sin \frac{ \Delta_{21} x }{2} 
\sin \frac{ \Delta_{31} x }{2}. 
\label{ampsq-interf:mu-e}
\end{eqnarray}
We note that the interference term, the second equation of
\eqref{ampsq-interf:mu-e}, displays the key feature of the
problem. That is, it consists of two terms: one that depends on
$\delta$ and another that does not. Therefore, observing effect of
$\delta$ is due to the quantum interference between the atmospheric
and the solar amplitudes, but only a part of the total effect.  A
claim of observation of the quantum interference between the
atmospheric and the solar amplitudes requires the observation of
\emph{both} terms in {\eqref{ampsq-interf:mu-e}} with the
\emph{correct} magnitudes; {\it i.e.}  a measurement of $\delta$ is
\emph{not} the same as a measurement of the interference effect.

Now, we discuss the $\nu_{e} \rightarrow \nu_{e}$ channel, which is identical to the $\bar\nu_{e} \rightarrow \bar\nu_{e}$ channel due to $CPT$-invariance.  The atmospheric and solar amplitudes are written as
\begin{eqnarray} 
&& 
S_{ee}^{ \text{atm} } =  
2 s^2_{13} 
e^{-i \frac{\pi}{2} } e^{ - i \frac{ \Delta_{31} x }{2} } 
\sin \frac{ \Delta_{31} x }{2}, 
\nonumber \\ 
&&
S_{ee}^{ \text{sol} } =  
2 s^2_{12} c^2_{13}
e^{-i \frac{\pi}{2} } e^{ - i \frac{ \Delta_{21} x }{2} } 
\sin \frac{ \Delta_{21} x }{2}.
\label{atm-sol-amplitude-ee}
\end{eqnarray}
Due to un-oscillated ``1'' in Eq.~\eqref{S-matrix-disapp}, the
$\nu_{e}$ survival probability $P(\nu_{e} \rightarrow \nu_{e})$ takes
a slightly complicated form, but can be written in a similar form as
in the appearance channel,
\begin{eqnarray} 
&&
P(\nu_{e} \rightarrow \nu_{e}) 
= P_{e e}^{ \text{non-int-fer} } 
+ P_{e e}^{ \text{int-fer} }, 
\label{probability-decomposed-ee} 
\end{eqnarray}
where 
\begin{eqnarray} 
&& 
P_{e e}^{ \text{non-int-fer} } \equiv 
1 + \vert S_{ee}^{ \text{atm} } \vert^2 + \vert S_{ee}^{ \text{sol} } \vert^2 
+ 2 \mbox{Re} \left[ S_{ee}^{ \text{atm} } + S_{ee}^{ \text{sol} } \right] 
\nonumber \\ 
&&=
1 - \sin^2 2\theta_{13} \sin^2 \frac{ \Delta_{31} x }{2} 
\nonumber \\
&&\hspace{10mm}
- 4 s^2_{12} c^2_{13} \left( 1 -  s^2_{12} c^2_{13}  \right) \sin^2 \frac{ \Delta_{21} x }{2}, 
\nonumber \\
&& 
P_{e e}^{ \text{int-fer} } \equiv 
2 \mbox{Re} \left[ \left( S_{ee}^{ \text{atm} } \right)^* S_{ee}^{ \text{sol} } \right] 
\nonumber \\ 
&&=
2 \sin^2 2\theta_{13} 
s^2_{12} 
\sin \frac{ \Delta_{31} x }{2} 
\cos \frac{ \Delta_{32} x }{2}
\sin \frac{ \Delta_{21} x }{2}.
\label{ampsq-interf:e-e}
\end{eqnarray}

\section{How to observe the quantum interference effect}
\label{sec:observation}\label{sec:JUNO}

We briefly discuss how to pin down the quantum interference effect
between the atmospheric and solar amplitudes. Once we obtain the
expression of the oscillation probability as
\begin{eqnarray} 
&&
P(\nu_{\beta} \rightarrow \nu_{\alpha}) 
= P_{\beta \alpha}^{ \text{non-int-fer} }
+ P_{\beta \alpha}^{ \text{int-fer} }, 
\label{probability-decomposed-ba} 
\end{eqnarray}
we can define a ``test oscillation probability'' by introducing the
$q$ parameter as
\begin{eqnarray}
&& P(\nu_\beta \rightarrow \nu_\alpha) = 
P_{\beta \alpha}^{ \text{non-int-fer} }
+ q P_{\beta \alpha}^{ \text{int-fer} }. 
\label{ansatz}
\end{eqnarray}
By fitting the data with the test oscillation probability
\eqref{ansatz}, we would obtain 1-dimensional $\chi^2$ (1 DOF) for the
$q$ parameter. We note that, in the case of appearance experiments, we
marginalize over $\delta$ as well as the other mixing parameters in
the experimentally allowed ranges.

Though our discussion in this paper covers both the appearance and the
disappearance experiments in vacuum, the analysis of the appearance
channel in accelerator LBL experiments requires treatment of the
matter effect~\cite{HMP-matter}, which is beyond the scope of this paper. 

The experimental setting of JUNO \cite{An:2015jdp} is {\em uniquely}
suited for our purpose of observing the interference effect between
the atmospheric and solar oscillations. In JUNO, the solar and
atmospheric oscillation effects coexist with their full magnitudes at
the same detector. Both oscillations are fully developed and have left
the linear regime of $\sin \frac{\Delta_{k1} x}{2}$ . Even though the
atmospheric oscillation may be small wiggles over the long-wavelength
solar oscillation, the very good energy resolution of the JUNO
detector aims at its precision measurement. Therefore, JUNO is an
ideal experiment for the purpose of detecting the atmospheric - solar
interference effect. It is very likely the best choice among all
possible experiments, ongoing or planned, in vacuum and in matter.

\begin{figure}[t]
\vglue 0.2cm
\begin{center}
\includegraphics[width=\columnwidth]{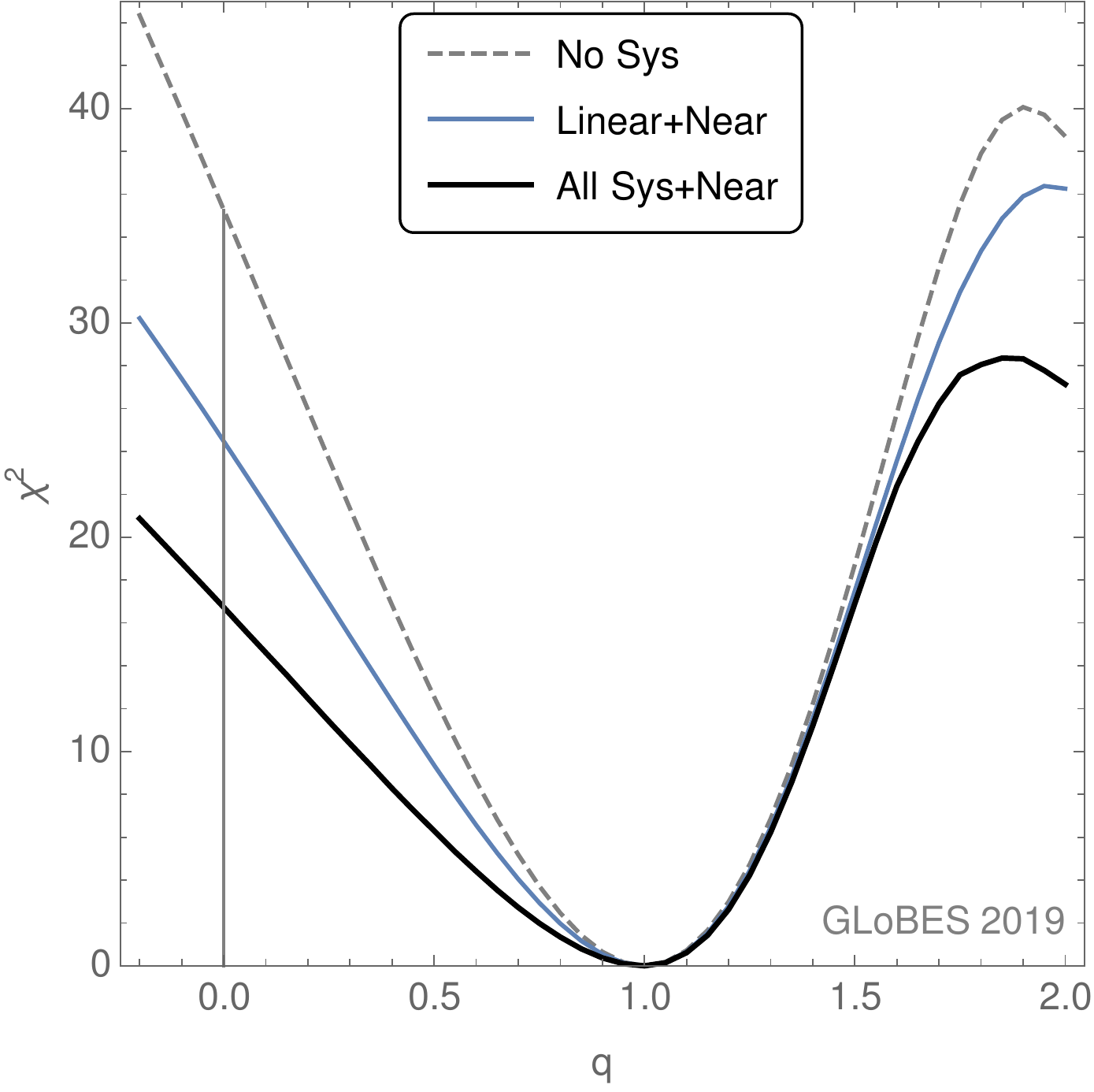}
\end{center}
\vglue -0.6cm
\caption{$\chi^2$ as a function of the $q$ parameter introduced in
  Eq. \eqref{ansatz}. The thick solid line includes all systematics
  with a near detector, whereas the blue line assumes the energy scale
  errors are linear. The gray dashed line is computed without any
  systematics and with the assumption of a perfect knowledge of the
  initial neutrino flux and no near detector.}
\label{chi2-curve}
\end{figure}

Here, we describe in detail the procedure of our statistical
analysis. Using GLoBES \cite{Huber:2005,Huber:2007}, we set up an
experiment with two detectors: a JUNO-like far detector with a
fiducial mass of $20\,\text{kt}$ and an energy resolution of
$3\%/\sqrt{E}$ at a distance of $53\,\text{km}$ from a nuclear reactor
source with a total power of $36\,\text{GWth}$, and a TAO-like
\cite{deAndre:2019} near detector with a fiducial mass of
$1\,\text{ton}$ and an energy resolution of $1.7\%/\sqrt{E}$ at a
distance of $30\,\text{m}$ from a $4.6\,\text{GWth}$ nuclear reactor
core; we assume a total data taking time of 6 years. For each
detector, we use a model for non-linear effects in the reconstruction
of the positron energy like that described in
Ref. \cite{Forero:2017vrg} up to cubic terms. To account for the
uncertainties in the reactor antineutrino flux prediction, we
conservatively introduce a nuisance parameter to each of our 100
energy bins with the spectrum computed before applying the energy
resolution function. This is equivalent to the assumption of \emph{no}
prior knowledge of fluxes, as in Ref. \cite{Forero:2017vrg}. For the
purposes of producing simulated data, we assume the normal ordering 
to be the true mass ordering and the relevant oscillation parameters to be
$\Delta m^2_{21}=7.54\times 10^{-5}\,\text{eV}^2, \Delta
m^2_{31}=2.43\times
10^{-3}\,\text{eV}^2,\theta_{12}=33.6^{\circ},\text{ and
}\theta_{13}=8.9^{\circ}$. For the analysis of the resulting data, we
fit the data obtained from the oscillation probability in
Eq.~\eqref{ampsq-interf:e-e} with that obtained using the oscillation
probability modified with the parameter $q$, as in Eq. \eqref{ansatz},
by minimizing the following $\chi^2$ function for various values of
$q$ while allowing all nuisance and standard oscillation parameters to
vary:
\begin{eqnarray}
&&
\chi^2 = \sum_{i,I}\frac{\qty(\phi_{\text{true},i}^I-\phi_{\text{fit},i}^I)^2}{\phi_{\text{true},i}^I}+\text{pull terms}\,\text{,}
\label{chi2}
\end{eqnarray}
where $\phi_{\text{true},i}^I$ and $\phi_{\text{fit},i}^I$ are the
simulated rate and modified rate, respectively, in the $i^\text{th}$
energy bin for the detector specified by $I=\text{Near,Far}$. The
``pull terms," defined in Eq.~\eqref{chi2pull}, provide a penalty for
$\theta_{13}$ with an
uncertainty of $\sigma_{\theta_{13}}=10\%$ and the nuisance
parameters $n_k$ for which uncertainties are $\sigma_k$:

\begin{eqnarray}
&&
\text{pull terms} = \frac{\qty(\theta_{13,\text{true}}-\theta_{13,\text{fit}})^2}{\sigma_{\theta_{13}}^2}+\sum_{k}\frac{n_k^2}{\sigma_k^2}\,\text{.}
\label{chi2pull}
\end{eqnarray}
The nuisance parameters included in the ``pull terms" encode the uncertainties for energy calibration (only linear terms), fiducial mass of each detector, and flux, as described in detail in Ref.~\cite{Forero:2017vrg}.

The resulting $\chi^2$ curve is shown as the thick black line in
figure \ref{chi2-curve}.  At $q=0$, the value of $\chi^2$ is 16.7, so the
interference effect would be able to be seen in JUNO with a significance
of more than $4\sigma$.

The same analysis procedure is repeated except assuming that the
energy calibration error for each detector is linear (blue solid
line), and then without a near detector while assuming perfect knowledge of
detector and source systematics (gray dashed line).

Note that there is a potential model-dependence, in that we assume that atmospheric oscillation experiments observe $\Delta m_{31}^2$.  If, instead, we assume that they measure $\Delta m_{32}^2$, the value of $\chi^2$ at $q=0$ for the scenario with a near detector and most conservative systematics is still 16.7.

\section{Interference effect and Mass ordering}

It is a natural question to ask how the sensitivity to the
interference effect depends on the neutrino mass ordering, and
conversely, whether the capability to determine the mass ordering is
due to the interference between the atmospheric and solar
waves. Hereafter, we use the abbreviation ``NO'' and ``IO'' for the
normal and the inverted orderings, respectively.

\begin{figure}[t]
\vglue 0.2cm
\begin{center}
\includegraphics[width=\columnwidth]{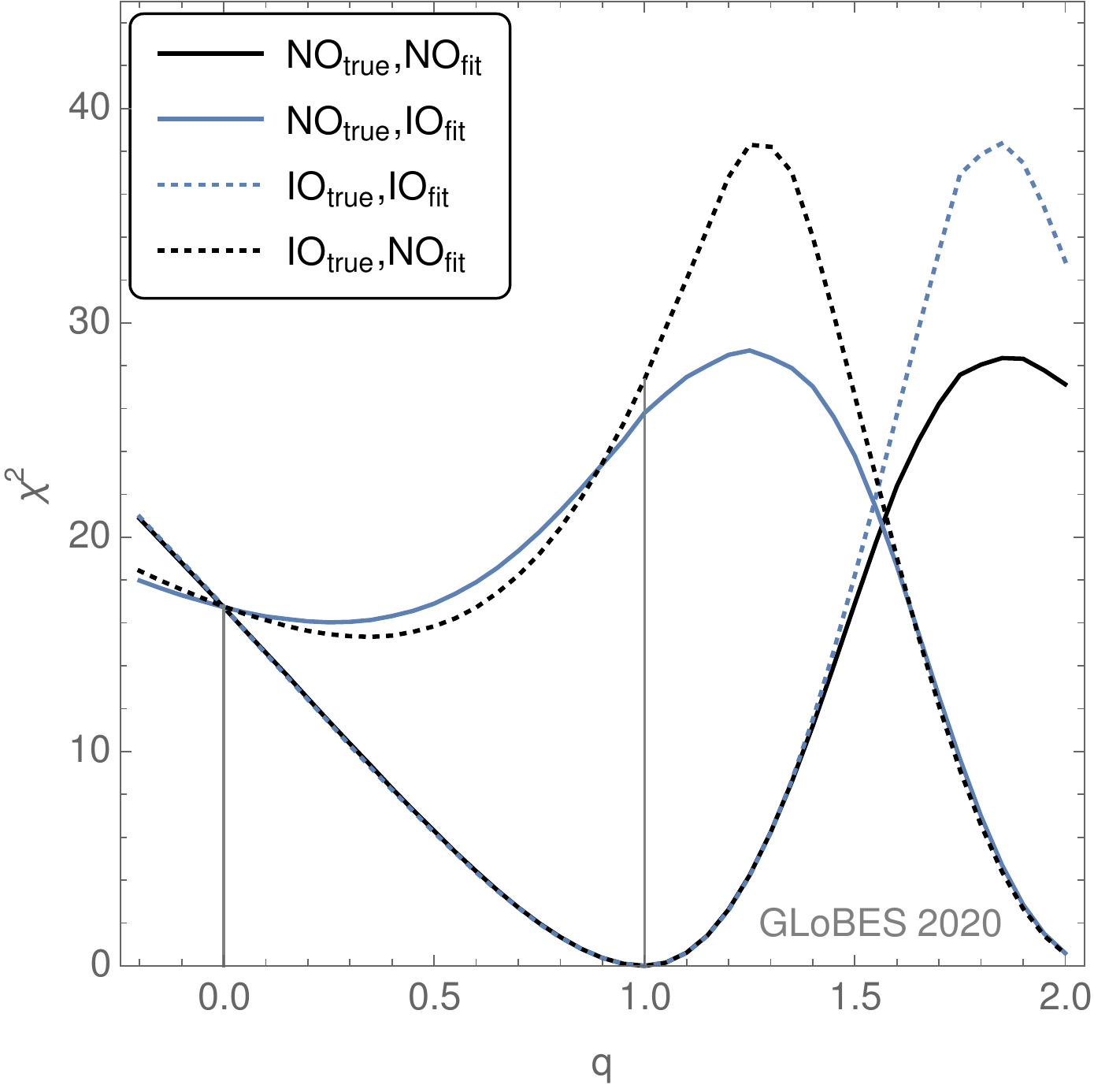}
\end{center}
\vglue -0.6cm
\caption{$\chi^2$ as a function of the $q$ parameter defined in Eq. \eqref{ansatz} with all the systematics and with a near detector. The solid lines are for the case that the true mass ordering is normal, which are fit with the hypotheses of the NO (black line) or IO (blue line). The dotted lines are for the case of true IO, which are fit with the NO (black line) or IO (blue line) hypotheses. The black solid curve is identical with the one in figure \ref{chi2-curve}. 
}
\label{Int-MO}
\end{figure}

We start by recalling that the interference term, the second term in
eq.~\eqref{ampsq-interf:e-e}, must be the origin of sensitivity to the
mass ordering because it is the only term which is odd under the inversion NO $\leftrightarrow$ IO. 
This seems to support the notion expressed in
Refs.~\cite{Petcov:2001sy,Choubey:2003qx} that the mass ordering
resolution is due to the interference between the atmospheric - solar
waves. However, we will show that the reality is a little more complicated.\footnote{
Note also, that the interference term defined in Refs.~\cite{Petcov:2001sy,Choubey:2003qx} is different from ours. 
}

The mass ordering dependence of $\chi^2 (q)$ is examined in
Fig.~\ref{Int-MO} using the most conservative systematics and including
a near detector, as in the previous section. If we use the NO (IO) for
both the true mass ordering and the test probability for fitting, the
black solid (blue dotted) curve results. These two curves indicate that
sensitivity to the interference does not depend on the mass ordering.

The behavior of $\chi^2 (q)$ drastically changes if the JUNO data is
fit with the wrong mass ordering. In Fig.~\ref{Int-MO}, the
black dotted (blue solid) curve is the case of true NO (IO)
fit with the IO (NO) hypothesis. The large value of $\chi^2 (q) \approx
25$ at $q=1$ tells us that JUNO can refute the wrong mass ordering at
a confidence level around $5\sigma$ in both cases of the true mass
orderings. This result is consistent with the one in
Ref.~\cite{Forero:2017vrg}. The precise value depends on the details
of the systematics implementation, see also Ref.~\cite{An:2015jdp} and
the references cited therein, but is not germane.

One might naively expect that detection of the interference
term~\eqref{ampsq-interf:e-e} would be trivial for JUNO, as it is large,
$\sim \sin^2 2\theta_{13} \simeq 0.1$. However, this sensitivity is
being reduced by a combination of not knowing the values of
the oscillation parameters precisely enough to start with and
cancellation occurring due to the energy bins. In fact, one can show
analytically that an integration over a $\frac{1}{4}$ period of the
atmospheric-scale oscillation of $P_{e e}^{ \text{int-fer} }$ in
\eqref{ampsq-interf:e-e} cancels the contribution from the adjacent
$\frac{1}{4}$ period under the approximation $\sin \Delta_{21} x \sim
\Delta_{21} x$. Then, it can be translated into the cancellation among
different energy bins in integration over $1/E$, which leads to an
imperfect but efficient cancellation in energy space. It is
conceivable that such cancellation contributes substantially to the
behavior of $\chi^2 (q)$ in the right and wrong mass-ordering fits.

Another key feature of the question is the minimum of $\chi^2 (q)$
at $q \simeq 2$ in the wrong mass-ordering fit. It is produced by
allowing both $\theta_{12}$ and $\Delta m^2_{31}$ to float freely in
the fit. If we instead kept all the oscillation parameters fixed, we
would find that all of the curves become parabolic with their
minimum at $q\simeq1$, albeit with the wrong mass-ordering fit curves
having a much larger value at the minimum. The minimum around $q=2$
arises as a combined effect of shifting $\Delta m^2_{31}$ by about 1\%
from its input value and $\sin^2\theta_{12}$ by about 3\%,
respectively. The occurrence of this minimum is independent of
systematics and energy resolution, which can be demonstrated by using
a pseudo-$\chi^2$, $\hat\chi^2$, defined as
\begin{equation}
  \hat\chi^2(q)=\int\,dE\, \left[P_{NO}(E,q=1)-P_{IO}(E,q)\right]^2\,.
  \label{pseudo-chi2}
  \end{equation}
Apart from an overall scale factor, $\hat\chi^2$ reproduces the
q-dependence shown Fig.~\ref{Int-MO} if $\theta_{12}$ and $\Delta
m^2_{31}$ are allowed to float. Moreover, we find that the value of
$q$ for which the IO minimum occurs scales like
$\sin^{-2}\theta_{12}$. We conclude that this second minimum is due to
a complete cancellation at the probability level by accident, {\it
  i.e.} the values of oscillation parameters it occurs at do not
represent an intrinsic symmetry of the oscillation probability. This
rather striking behavior also demonstrates that a determination of the
mass ordering, which entirely takes place at $q=1$, is \emph{not
  equivalent} to study the question of whether the q-term ({\it i.e.},
the interference term) is present at all.

The origin of the local deep minimum of $\chi^2 (q)$ at $q \sim
2$ can also be understood by following the analysis technique outlined
in Ref.~\cite{Learned:2006wy}, which is using the Fourier transform of
the event spectrum or probability as a function of
$L/E$.\footnote{Note, that the Fourier approach, while conceptually
  very clear, is not well suited to a full study including systematic
  effects on the energy scale and thus, is not used for actual
  sensitivity estimates, see for example Ref.~\cite{An:2015jdp}. }
In the Fourier spectrum, one observes a main peak at $\Delta m_{31}^2$
with a shoulder at $\Delta m_{31}^2\pm\Delta m^2_{21}$ and the mass
hierarchy is determined by the \emph{relative} position of shoulder
and main peak. Note that the absolute positions of the peaks are only
known within the uncertainty of $\Delta m^2_{31}$, which is much larger
than their separation.

Varying $q$ changes the relative amplitudes of the peaks and can thus
lead to a confusion of mass ordering. If we use $\hat\chi^2$ in
eq.~\eqref{pseudo-chi2} with the true NO, which has a local minimum
at $q\simeq 2$, the relative positions of the higher and lower
peaks exchange positions when $q$ is increased from $q=1$ and
$q\simeq2$.

Conceptually, it is possible to imagine a world in which $\Delta
m_{31}^2$ is determined with superb accuracy ($\ll \Delta m_{21}^2$). In
this case, one can show numerically that the degeneracy indeed goes
away, and hence, there is no confusion, since now the position of the
peaks, instead of their relative heights, can be used to determine the
mass hierarchy. Thus, in that case, the sensitivity to mass ordering
exists in a robust way independent of the potential strength
of the interference term.

To summarize, we failed to see evidence for the intimate, direct
connection between sensitivity to the mass ordering and the
atmospheric - solar interference effect.

\section{Summary}

In this paper, we have shown that, in vacuum, a natural and general way of decomposing the oscillation amplitudes into solar and atmospheric parts is possible for appearance and disappearance channels. This decomposition is exact and relying neither on the hierarchical values of the two $\Delta m^2$ nor on the actual values of observed oscillation parameters. With this amplitude decomposition, it becomes possible to define the effect of interference between the two partial amplitudes.  For appearance channels, the interference term contains the $CP$-phase $\delta$, but also terms independent of it.

In the $\bar{\nu}_{e}$ ($\nu_{e}$) disappearance channel, the oscillation amplitude does \emph{not} depend on $\delta$, and hence, the interference effect we saw has nothing to do with the $CP$-phase. The nature of the interference phenomena indicated by these features is a dynamical, quantum mechanical interference inside the three family of neutrinos, not particularly related to the $CP$-violating phase. We show, by detailed numerical calculation, that JUNO can observe this interference effect with more than $4\,\sigma$ significance.

We have also discussed the relationship between the interference effect and sensitivity to the mass ordering resolution. We argued that though the latter comes from the interference term in eq.~\eqref{ampsq-interf:e-e}, we see no supporting evidence for the hypothesis of the mass ordering resolution being equivalent to the atmospheric - solar wave interference effect.

\begin{acknowledgments}
One of the authors (H.M.) thanks Takaaki Kajita and Hiroshi Nunokawa
for intriguing conversations while this project was still in its
infancy. The work of P.H. and R.P. is supported by the US Department of
Energy Office of Science under award number \protect{DE-SC0020262}.

\end{acknowledgments}

Note added: After we completed this work, we have learned that the author of ref.~\cite{Bilenky:2012zp} used unitarity as in (4) to derive the alternative form of the oscillation probability in N flavor case which agrees with our formula for N=3. However, neither physics of atmospheric - solar interference, nor the amplitude decomposition with completeness is discussed in that paper.

\end{document}